\documentclass[12pt]{article}
\usepackage[a4paper,margin=2.2cm]{geometry}
\usepackage{rotating}
\usepackage{epsf}
\usepackage{graphics}

\title{Tracker coupled quintessence}
\author{\\F. Shojai, R. Moti and F. Najdat \\
Department of Physics, University of Tehran, Tehran, Iran.} 
\date{}
\usepackage{amsmath}
\usepackage{graphicx}
\begin{document}
\maketitle
\begin{abstract}
In this paper two tracker solutions for conformally coupled quintessence model is obtained. The first solution is determined by dividing the evolution of the universe to matter and dark energy dominated epochs and then the allowed bound on the equation of state parameter is obtained, such that a tracker behavior is resulted. For the other solution, the quintessence potential is assumed to be  a linear combination of two exponential functions. Then we have found an exact solution which can be tracker by adjusting some parameters.    
\end{abstract}
\section{Introduction}
Brans--Dicke scalar-tensor theory \cite{brans} is one of the simplest scalar-tensor theories motivated by Mach's principle. To satisfy this principle, Brans and Dicke assume that the gravitational constant is not really a constant but it is a scalar field ($1/\phi)$. Adopting the Einstein--Hilbert action with this assumption, one arrives at a non--minimal coupling of $\phi$ and gravity. This represents the Jordan picture of Brans--Dicke theory where the scalar field does not appear in the matter field sector. As a result the matter fields do not feel any force from the scalar field and the equivalence principle holds. 

It is clear that this theory is basically different from general relativity since the gravitational field is described not only by the metric but also by the scalar field. Moreover there is a dimensionless coupling constant which can be bounded by observational data. According to various observations, this parameter should be very large \cite{will}. There is a standard belief that in the limit $\omega\rightarrow\infty$, this theory is not distinguishable from general relativity \cite{faraoni}. But this is not satisfied generally. In fact there are some solutions of Brans-Dicke theory which doesn't lead to a corresponding solution of general relativity in this limit \cite{bane}. 

By a conformal transformation, one can switch to the Einstein picture in which the scalar field is conformally coupled to matter field. This means that the metric in the matter action in the Jordan picture is  replaced with the metric times an exponential factor of the      
scalar field in the Einstein picture.
In this picture the equivalence principle is broken and an extra force is appeared in the right hand side of the geodesic equation. Depending on the potential of the scalar field, it can be shown that the mass of the scalar field depends on the local mass density. This is called chameleon mechanism \cite{cham}. This effect suppresses the influence of extra force in the solar system experiments.

 Recently scalar tensor theories have been attracted interest  in studying the low energy limit of other theories and in providing  a suitable candidate for dark energy component of the universe. The first one provides an inflation epoch and reheating period in the early universe and the latter is what we have focused on here.
 
Describing the accelerated expansion of the universe by vacuum energy (cosmological constant) suffers from some conceptual problems \cite{wein}, like: Why the estimated vacuum energy density is about 120 orders of magnitude smaller than what is predicted by quantum field theory (fine tuning problem)? And why the densities of dark energy and dark matter are comparable at the present epoch  (coincidence problem)? One of the natural solutions is to use a scalar field as a dynamical source of dark energy, named quintessence \cite{sean}.

Following this, here we are interested to use Brans-Dicke theory in the Einstein picture as a source of dark energy \cite{conformal, roshan}. 
The phase space analysis of this theory in the Einstein\rq{}s picture is discussed in \cite{Holden} assuming an exponential potential.
In such a model the scalar field has a conformally coupled interaction with matter component, for other interaction terms see \cite{interaction}. For the late time evolution of the universe, we assume that the content of the universe is two fluids, non-relativistic matter and dark energy.   

The outline of this paper is as follows.  Starting from Brans-Dicke action in section 2, we shall find some new cosmological solutions for spatially flat universe for both dark matter dominated  
and dark energy dominated epochs in section 3. To obtain these, we assume a constant equation of state for the scalar field but different in each epoch. As we shall see, this leads to a tracker potential which is not in exponential form as it is in the case of non-interacting quintessence.  
The validity of the solutions in each epoch and the tracker behavior of dark energy component in the matter dominated epoch lead to some bounds on the equation of state parameter.  The behavior of scale factor, the matter--energy density ratio, the deceleration parameter and the form of potential at any epoch are obtained in section 4.

Finally assuming a linear combination of two exponential functions for the potential, we obtained  an exact tracker solution for interacting Brans-Dicke quintessence. This solution is valid for both the matter and dark energy dominated epochs. At the end, we present the general solution of equations for a given parametrized form of equation of state.

\section{Interacting Brans-Dicke quintessence}
The action of Brans-Dicke theory in the Einstein picture is:
\begin{equation}
S_E=\int d^4x\sqrt{-g}\left [\frac{R}{12\alpha^2}-\frac{1}{2}(\nabla\phi)^2-V(\phi) + {\cal L}_m(e^{-2\alpha\phi}g_{\mu\nu}, \phi_m)\right ]
\end{equation}
where $\alpha=\sqrt{\frac{4 \pi G}{3}}$. This leads to the following field equations for a spatially flat Friedmann-Robertson-Walker universe:
\begin{equation}
H^2=2\alpha^2(\rho_\phi+\rho_m)
\label{1}
\end{equation}
\begin{equation}
\dot\rho_m+3H\rho_m=-\alpha\dot{\phi}\rho_m
\label{rohm}
\end{equation}
\begin{equation}
\dot{\rho_\phi}+3H(1+\omega)\rho_\phi=\alpha\dot{\phi}\rho_m
\label{2}
\end{equation}
where $\rho_{m}$ is the non-relativistic matter density and $\rho_{\phi}=\frac{1}{2}\dot{\phi}^2+V$. The scalar field pressure is defined as $p_{\phi}=\frac{1}{2}\dot{\phi}^2-V$  and $\omega=\frac{p_{\phi}}{\rho_{\phi}}$ is the equation of state parameter. Defining the dimensionless density parameters as:
\begin{equation}
 X_i=\frac{2\alpha^2\rho_i}{H_*^2}
\end{equation}
where we have used $H_*$ as the Hubble parameter at the time that the scalar field and matter densities are equal. We thus have:
\begin{equation}
\dot{\phi}=\frac{\beta H_*}{\sqrt{2}\alpha}\sqrt{ X_\phi}
\label{dotphi}
\end{equation} 
in which $\beta=\sqrt{1+\omega}$ and here we assume that it is constant in time.
Notice that since this model is a quintessence model, so $0\leq\beta^2\leq2$. 

It is convenient to rewrite equations of motion in terms of the above parameters:
\begin{equation}
H^2=H_*^2( X_\phi+ X_m)
\label{H}
\end{equation}
\begin{equation}
\dot{ X_m}+3H X_m=\frac{-\beta H_*}{\sqrt{2}}\sqrt{ X_\phi} X_m
\label{omegam}
\end{equation}
\begin{equation}
\dot{ X_\phi}+3H\beta^2 X_\phi=\frac{\beta H_*}{\sqrt{2}}\sqrt{ X_\phi} X_m
\label{omegaphi}
\end{equation}
\section{The dynamics}
\subsection{Dark matter dominated epoch}
In the matter dominated epoch, the scalar field density is much less than the matter density. Ignoring the
first term in equation (\ref{H}), substituting it into (\ref{omegam}) and (\ref{omegaphi}) and denoting the equation of state of quintessence for this case as $\tilde{\omega}$ , we find: 
\begin{equation}
2(\sqrt{ X_m})^\prime +3\sqrt{ X_m}=\frac{-\tilde{\beta}}{\sqrt{2}}\sqrt{ X_\phi}
\label{omegam1}
\end{equation}
and
\begin{equation}
2({\sqrt{ X_\phi})^\prime +3\tilde{\beta}}^2\sqrt{ X_\phi}=\frac{\tilde{\beta}}{\sqrt{2}}\sqrt{ X_m}
\label{omegaphi1}
\end{equation}
where a prime denotes the derivative with respect to the logarithm of the normalized scale factor $N=\ln{\frac{a}{a_*}}$.
Remembering that the initial condition is: $ X_\phi(N_*=0)= X_m(N_*=0)=1/2$. For constant $\tilde{\omega}$, (\ref{omegam1}) and (\ref{omegaphi1}), are two coupled first order differential equations  which lead to the following analytical solutions for  $ X_\phi$ and $ X_m$:

\begin{itemize}
\item {(a) If $0.628<\tilde{\beta}^2<1.595$}:
\begin{equation}
 X_\phi(N)=\frac{e^{2\mu N}}{2\lambda}\left[ \sqrt{\lambda}\cos\frac{\sqrt{\lambda}}{4} N+\left(\tilde{\beta}\sqrt{2}+3(1-\tilde{\beta}^2)\right)\sin\frac{\sqrt{\lambda}}{4} N\right]^2 
\label{solxmsin}
\end{equation}
\begin{equation}
 X_m(N)=\frac{e^{2\mu N}}{2\lambda}\left[ \sqrt{\lambda}\cos\frac{\sqrt{\lambda}}{4} N-\left(\tilde{\beta}\sqrt{2}+3(1-\tilde{\beta}^2)\right)\sin\frac{\sqrt{\lambda}}{4} N\right]^2
\label{solymsin}
\end{equation}

\item{(b) If $0<\tilde{\beta}^2<0.628$ or $1.595<\tilde{\beta}^2<2$}:
\begin{equation}
 X_\phi(N)=-\frac{e^{2\mu N}}{2\lambda}\left[ \sqrt{-\lambda}\cosh\frac{\sqrt{-\lambda}}{4} N+\left(\tilde{\beta}\sqrt{2}+3(1-\tilde{\beta}^2)\right)\sinh\frac{\sqrt{-\lambda}}{4} N\right]^2 
\label{solxmsinh}
\end{equation}
\begin{equation}
 X_m(N)=-\frac{e^{2\mu N}}{2\lambda}\left[ \sqrt{-\lambda}\cosh\frac{\sqrt{-\lambda}}{4} N-\left(\tilde{\beta}\sqrt{2}+3(1-\tilde{\beta}^2)\right)\sinh\frac{\sqrt{-\lambda}}{4} N\right]^2 
\label{solymsinh}
\end{equation}
\end{itemize}
where $\mu=-\frac{3(1+\tilde{\beta}^2)}{4}$ and $\lambda=38\tilde{\beta}^2-9(1+\tilde{\beta}^2)^2$.
The bounds of $\tilde\beta^2$, i.e. the numbers $0.628$ and $1.595$, are the roots of the secular of the differential equations (\ref{omegam1}) and (\ref{omegaphi1}).
The above solutions should lead to $ X_m\gg X_\phi$. To check this, note that $N$ is negative in the matter dominated era. Setting  $z=1000$ for the beginning of matter dominated universe, we get the following bound for $\tilde{\beta}$:      
\begin{itemize}
\item {(a)}
\begin{equation}
 X_m\gg X_\phi\Longrightarrow \left(\tilde{\beta}\sqrt{2}+3(1-\tilde{\beta}^2)\right)\sin{\frac{\sqrt{\lambda}}{2} N}\ll 0
\end{equation}
This leads to:
\begin{equation}
 0.628<\tilde{\beta}^2<0.757\  \text{or}\  1.466<\tilde{\beta}^2< 1.595
 \label{bo1}
\end{equation}
\item {(b)}
\begin{equation}
 X_m\gg X_\phi \Longrightarrow \left(\tilde{\beta}\sqrt{2}+3(1-\tilde{\beta}^2)\right)\sinh{\frac{\sqrt{\lambda}}{2} N}\ll 0
\end{equation}
Therefore:
\begin{equation}
0 < \tilde{\beta}^2< 0.628
\label{bo2}
\end{equation}
\end{itemize}
Since $2\mu$ is always larger than $\sqrt{|\lambda|}/4$, the scale of variation of terms in the brackets in equations 
(\ref{solxmsin})--(\ref{solymsinh}), is larger than that of the exponential terms. This suggests that these solutions exhibit a tracking behavior. It is known that under tracking conditions
 $\rho_\phi$ must be less than $\rho_m$ for some range of initial conditions (for the necessary conditions of having tracker solution see \cite{zlatev} for non-interacting quintessence and \cite{roshan} for Brans-Dicke type interaction).
 Therefore during tracking period it is required that $\mid X_\phi^{'}\mid<\mid X_m^{'}\mid$. Using a simple calculation one can show that this condition is satisfied for the range of $\tilde{\beta}$ given above.  

In order to discuss the convergence to the tracker solution, (i.e. to check if the solution of matter dominated era is attractor or not), we apply a small perturbation $\delta X_\phi$ and $\delta X_m$ to the solutions (\ref{solxmsin})-(\ref{solymsinh}). As a result  the following equation for $\delta X_\phi$ is obtained:
\begin{equation}
(\delta X_\phi)^\prime + 3(\delta X_\phi) = \frac{\tilde{\beta}}{\sqrt 2}\left(\sqrt {\frac{ X_\phi}{ X_m}}(\delta X_m)+\frac{1}{2}\sqrt {\frac{ X_m}{ X_\phi}}(\delta X_\phi)\right)
\label{delta1}
\end{equation}
In the matter dominated era, $ X_\phi\ll X_m$,  we can ignore the first term in the above equation. This yields to a decoupled equation for $\delta X_\phi$, which can be easily solved for the two cases mentioned before:
\begin{itemize}
\item{(a)}
\begin{equation}
\delta X_\phi=\delta X^0_\phi e^{\frac{3(1-5\tilde{\beta}^2)}{4}N}\left[\sqrt{\lambda}\cos\frac{\sqrt{\lambda}}{4} N+\left(\tilde{\beta}\sqrt{2}+3(1-\tilde{\beta}^2)\right)\sin\frac{\sqrt{\lambda}}{4} N\right]^{-1}
\label{at1}
\end{equation}
\item{(b)}
\begin{equation}
\delta X_\phi=\delta X^0_\phi e^{\frac{3(1-5\tilde{\beta}^2)}{4}N}\left[ \sqrt{\lambda}\cosh\frac{\sqrt{\lambda}}{4} N+\left(\tilde{\beta}\sqrt{2}+3(1-\tilde{\beta}^2)\right)\sinh\frac{\sqrt{\lambda}}{4} N\right]^{-1}
\label{at2}
\end{equation}
\end{itemize}
In order to have attractor behavior, it is necessary that the fractional perturbation $\delta X_\phi / X_\phi$ decays as $t$ increases. Using  (\ref{solxmsin}) and (\ref{at1}), a straightforward calculation yields $\tilde{\beta}^2>2/3$ which provides a tighter bound on  $\tilde{\beta}$ (see (\ref{bo1})). A similar calculation for the relation (\ref{at2}) leads to the same constraint on  $\tilde{\beta}$ which violates (\ref{bo2}). Therefore the solution (b) is not an attractor tracker solution. 

\begin{figure}[htp]
\centering
\includegraphics[scale=0.6]{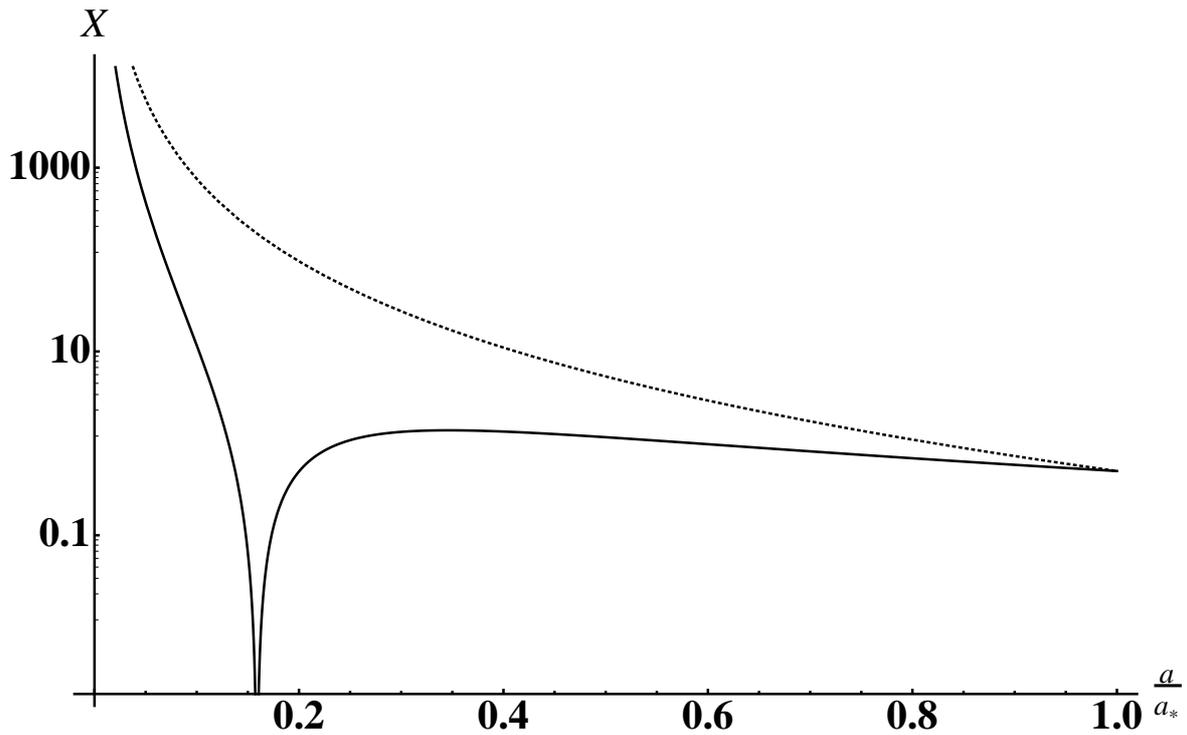}
\caption{ Evolution of dark matter (dotted line) and dark energy (solid line) in terms of normalized scale factor $a/a_*$ for dark matter epoch by taking $\tilde{\beta}=\sqrt{0.7}$.}
\label{f1}
\end{figure}

In summary  a tracker solution for Brans-Dicke type interacting quintessence is obtained with the following constraint:
\begin{equation}
-0.333<\tilde{\omega}<-0.243 \ \ \ \ \ \ \  \text{or}\  \ \ \ \ \ \  0.466<\tilde{\omega}<0.595
\end{equation}
Figure (\ref{f1}) shows the variation of dark energy and matter densities as a function of normalized scale factor in the matter dominated epoch. We see that the dark energy density first goes to zero (note the logarithmic scale in the plot), and after reaching to a local maximum, decreases. This is basically related to the interaction term, $\dot\phi\rho_m$. This local maximum has no significant role in the matter dominated epoch. One can easily see that at this maximum $X_\phi$ is very smaller than $X_m$ (For details see \cite{roshan}).
\subsection{Dark energy dominated epoch}
\begin{figure}[htp]
\centering
\includegraphics[scale=0.6]{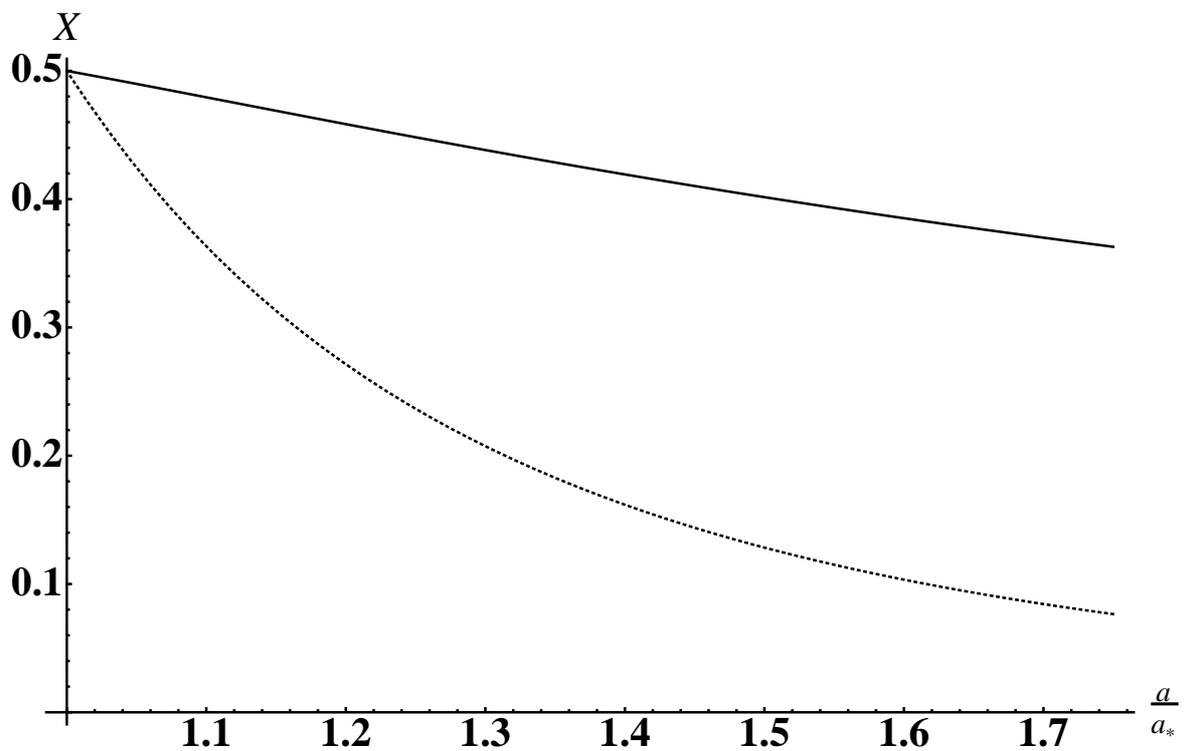}
\caption{ Evolution of dark matter (dotted line) and dark energy (solid line) in terms of normalized scale factor $a/a_*$ for dark energy epoch by taking $\bar{\beta}=0.5$.}
\label{f2}
\end{figure}

Once $ X_m$ falls well below $ X_\phi$, the second term in (\ref{H}) is negligible. This is the beginning of dark energy dominated epoch. Denoting the equation of state by $\bar{\omega}$, the conservation equations (\ref{omegam}) and (\ref{omegaphi}) can be written as follows:
\begin{equation}
 X_m^\prime +3 X_m =\frac{-\bar{\beta}}{\sqrt{2}} X_m
\label{omegam2}
\end{equation}
\begin{equation}
2 X_\phi^\prime +3\bar{\beta}^2 X_\phi=\frac{\bar{\beta}}{\sqrt{2}} X_m
\label{omegaphi2}
\end{equation}
with the solutions:
\begin{equation}
 X_\phi=\frac{1}{2}\left[\left(1-\frac{\bar{\beta}}{3\sqrt{2}(\bar{\beta}^2-1)-\bar{\beta}}\right)
e^{-3\bar{\beta}^2 N}+\frac{\bar{\beta}}{3\sqrt{2}(\bar{\beta}^2-1)-\bar{\beta}}
e^{-(3+\frac{\bar\beta}{\sqrt 2})N}\right]
\label{omegaphiphi}
\end{equation}
\begin{equation}
 X_m=\frac{1}{2}e^{-(3+\frac{\bar{\beta}}{\sqrt{2}}) N}
\label{omegamm}
\end{equation}
where  $ X_\phi(N_*)= X_m(N_*)=1/2$.
This solution is acceptable if the dark energy dominates and both densities are positive. A straightforward calculation shows that this does not imply any bound on $\bar{\beta}$.
The behavior of density parameters in this epoch is plotted in figure (\ref{f2}). 
\section{Evolution of the scale factor and potential}
\begin{figure}[htp]
\centering
\includegraphics[scale=0.6]{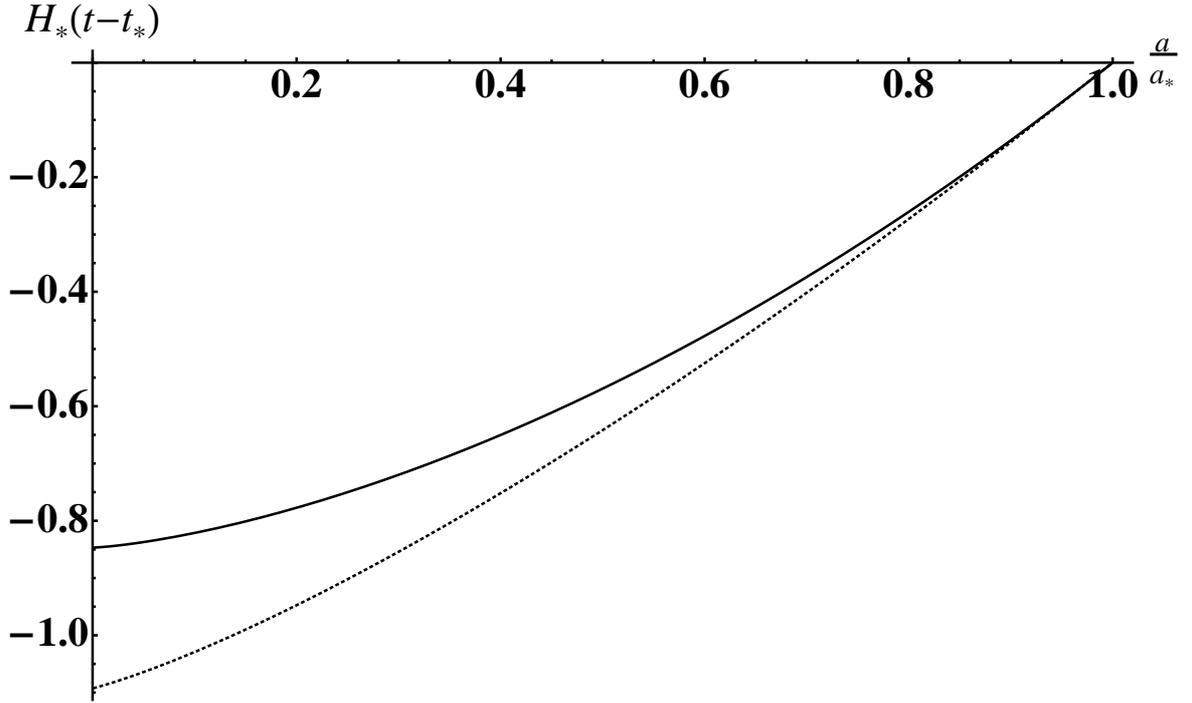}
\caption{ $t-t_*$ in units of Hubble time as a function of normalized scale factor $a/a_*$ for dark matter epoch.The solid and dotted lines correspond to $\tilde{\beta}=\sqrt{0.7}$ and $\tilde{\beta}=\sqrt{1.5}$}
\label{f3}
\end{figure}

\begin{figure}[htp]
\centering
\includegraphics[scale=0.6]{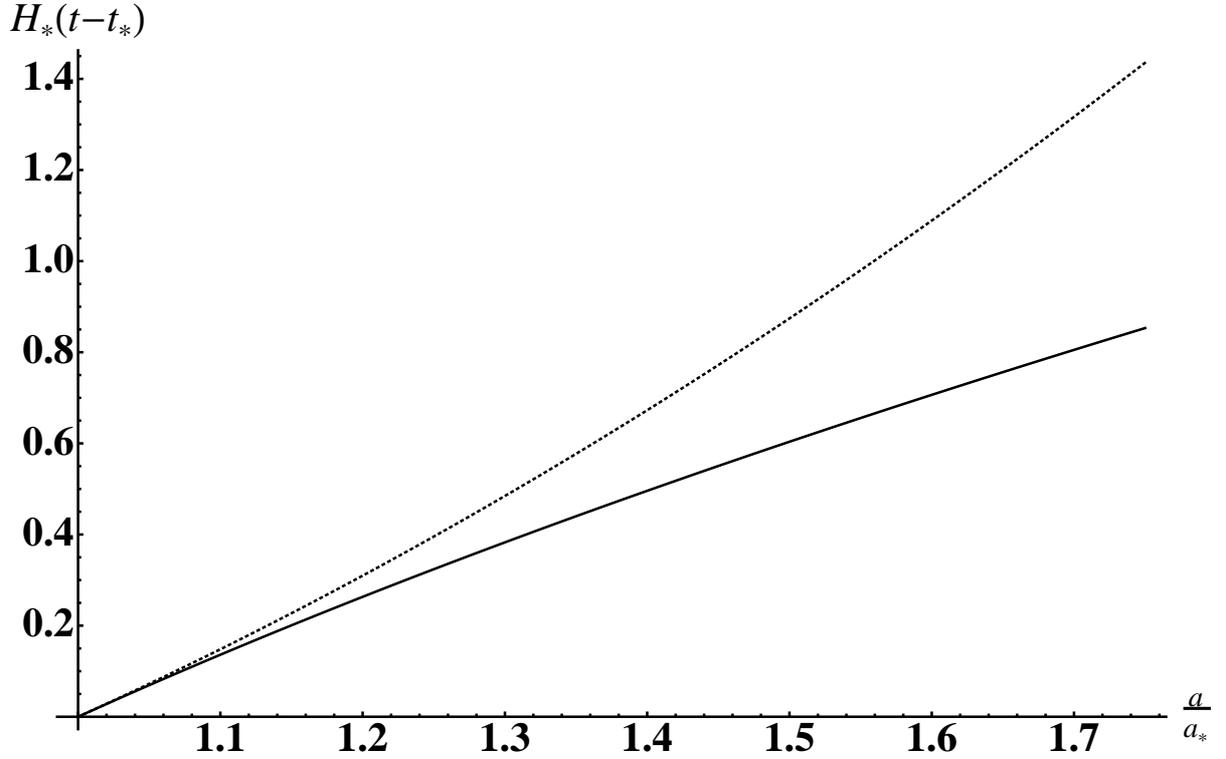}
\caption{$t-t_*$ in units of Hubble time as a function of normalized scale factor $a/a_*$ for dark energy epoch.The solid and dotted lines correspond to $\bar{\beta}=0.5$ and $\bar{\beta}=1.265$}
\label{f4}
\end{figure}

\begin{figure}[htp]
\centering
\includegraphics[scale=0.6]{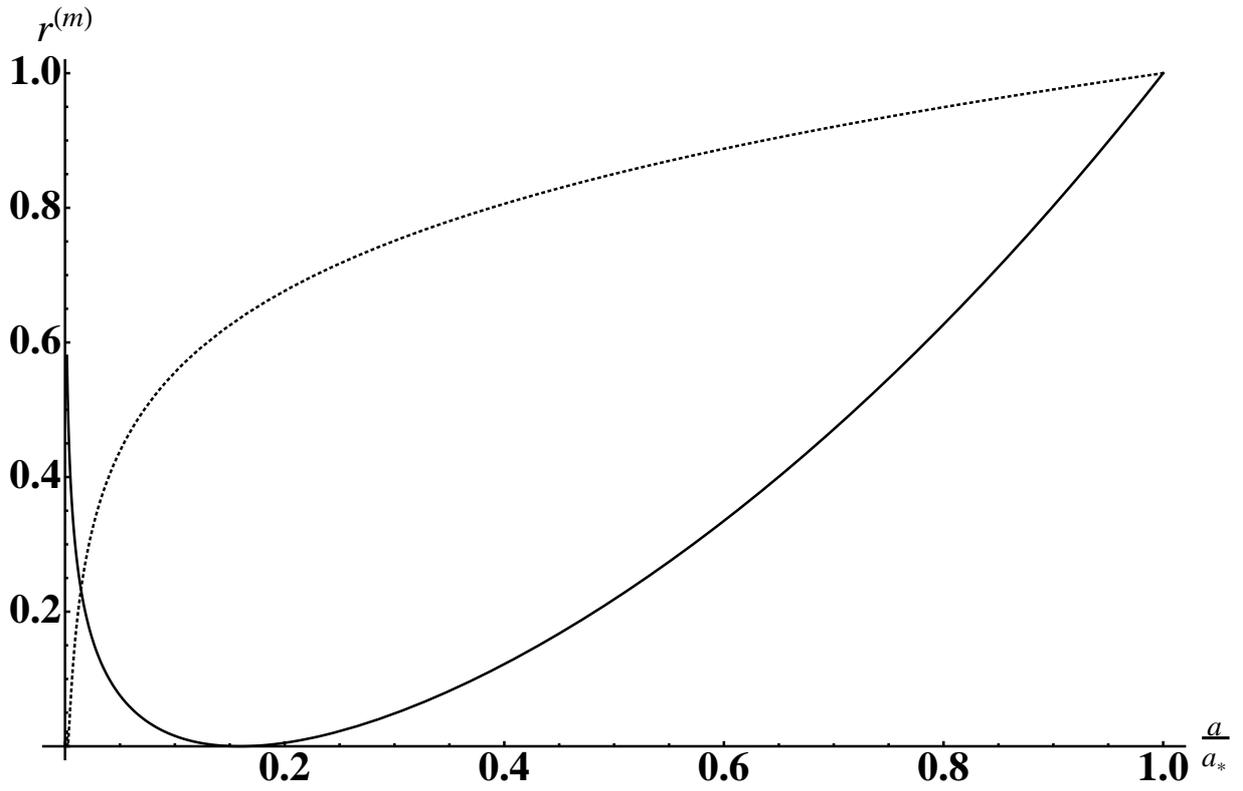}
\caption{Dark energy to dark matter ratio as a function of normalized scale factor $a/a_*$ for dark matter epoch. The solid and dotted lines correspond to $\tilde{\beta}=\sqrt{0.7}$ and $\tilde{\beta}=\sqrt{1.5}$}
\label{f5}
\end{figure}

\begin{figure}[htp]
\centering
\includegraphics[scale=0.6]{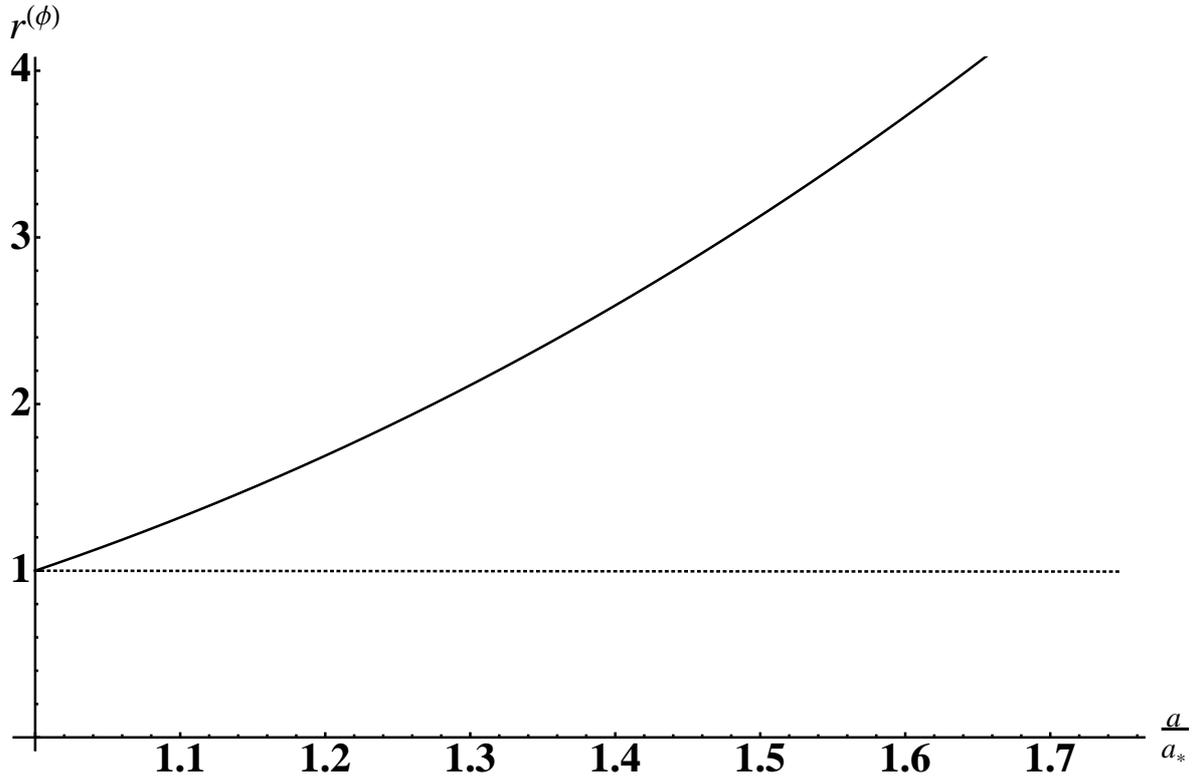}
\caption{Dark energy to dark matter ratio as a function of normalized scale factor $a/a_*$ for dark energy epoch. The solid and dotted lines correspond to $\bar{\beta}=0.5$ and $\bar{\beta}=1.265$ }
\label{f6}
\end{figure}

From equation (\ref{H}) and using the solutions (\ref{solxmsin})-(\ref{solymsin}) and (\ref{omegaphiphi})-(\ref{omegamm}), one can find the form of $N$ in terms of normalized cosmic time $H_* (t-t_*)$:
\begin{itemize}
\item{Matter dominated}
\begin{equation}
\frac{H_*(t-t_*)}{\sqrt{2\lambda}}=\int_0^N\frac{e^{-\mu N}dN}{\left[\sqrt\lambda\cos{\frac{\sqrt\lambda}{4} N}-\left(\tilde\beta\sqrt 2+3(1-\tilde\beta^2)\right)\sin{\frac{\sqrt\lambda}{4} N}\right]}
\label{n1}
\end{equation}
\item{Dark energy dominated}
\begin{equation}
H_*(t-t_*)=\int_0^N\frac{dN}{\sqrt{\frac{1}{2}\left[\left(1-\frac{\bar{\beta}}{3\sqrt{2}(\bar{\beta}^2-1)-\bar{\beta}}\right)
e^{-3\bar{\beta}^2 N}+\frac{\bar{\beta}}{3\sqrt{2}(\bar{\beta}^2-1)-\bar{\beta}}
e^{-(3+\frac{\bar\beta}{\sqrt 2})N}\right]}}
\label{n2}
\end{equation}
\end{itemize}
Figures (\ref{f3}) and (\ref{f4}) show plot of the relations (\ref{n1}) and (\ref{n2}) for various values of the equation of state. The evolution of dark energy to dark matter ratio (denoted $r^{(\phi)}$ and $r^{(m)}$ for dark energy and matter dominated era respectively) is shown in figures (\ref{f5}) and (\ref{f6}).

It is useful to express the deceleration parameter $q=-1-\frac{\dot H}{H^2}$ in terms of 
$r$. By a simple calculation one can show that $q^{(m)}=\frac{1}{2}(1+\tilde{\beta}\sqrt{\frac{r^{(m)}}{2}})$ which is positive as it is expected. A similar calculation yields: $q^{(\phi)}=-1+\frac{\bar\beta}{2}(3\bar\beta-\frac{1}{\sqrt 2 r^{(\phi)}})$ which shows that the cosmic acceleration starts at $r^{(\phi)}\mid_{{q^{(\phi)}=0}}=\frac{\bar\beta}{\sqrt 2(3\bar\beta^2-2)}$ and depends on the value of $\bar\beta$. Therefore the difference between $q^{(\phi)}$
and $-1$ (the deceleration parameter for a cosmological constant dominated universe), is a monotonic decreasing function which approaches to zero at late times.

It is interesting to find the form of scalar field potential that gives the solution (\ref{solxmsin})--(\ref{solymsin}) for matter and (\ref{omegaphiphi})--(\ref{omegamm}) for dark energy dominated universe. To find this, we write both $V$ and $\phi$ as functions of $N$. Using $V=(2-\beta^2)\rho_\phi/2$:
\begin{equation}
V(N)= \frac{(2-\beta^2) H_*^2  X_\phi(N)}{4\alpha^ 2}
\label{vv}
\end{equation}
and by integrating the continuity equation (\ref{rohm}):
\begin{equation}
\phi(N)= \phi_*-\frac{3N+\ln 2 X_m(N)}{\alpha}
\end{equation}
The parametric form of the scalar potential in different regimes is thus:
\begin{itemize}
\item{(a)}
\begin{equation}
\phi(N)= \phi_*-\frac{1}{\alpha}\left((2\mu+3)N+\ln{\frac{\left[\sqrt{\lambda}\cos\frac{\sqrt{\lambda}}{4} N-\left(\tilde{\beta}\sqrt{2}+3(1-\tilde{\beta}^2)\right)\sin\frac{\sqrt{\lambda}}{4} N\right]^2}{\lambda}}\right)
\label{phim}
\end{equation}
\begin{equation}
V(N)=\frac{(2-\tilde\beta^2)H_*^2 e^{2\mu N}}{8 \alpha ^2 \lambda}\left(\sqrt{\lambda}\cos\frac{\sqrt{\lambda}}{4} N+\left(\tilde{\beta}\sqrt{2}+3(1-\tilde{\beta}^2)\right)\sin\frac{\sqrt{\lambda}}{4} N\right)^2
\label{vm}
\end{equation}
\item{(b)}
\begin{equation}
\phi=\frac{\bar{\beta}}{\sqrt{2}\alpha}N
\label{phiphi}
\end{equation}
\begin{equation}
V(N)=\frac{(1-\omega)H_*^2}{8 \alpha ^2}\left[\left(1-\frac{\bar{\beta}}{3\sqrt{2}(\bar{\beta}^2-1)-\bar{\beta}}\right)
e^{-3\bar{\beta}^2 N}+\frac{\bar{\beta}}{3\sqrt{2}(\bar{\beta}^2-1)-\bar{\beta}}
e^{-(3+\frac{\bar\beta}{\sqrt 2})N}\right]
\label{vphi}
\end{equation}
\end{itemize}
By eliminating $N$ in the above equations, one can reconstruct the form of the potential. 
In the $\phi$ dominated case, a simple substitution from (\ref{phiphi}) in (\ref{vphi}) leads to:
\[
V(\phi)=\frac{(1-\omega)H_*^2}{8 \alpha ^2}\left[\left(1-\frac{\bar{\beta}}{3\sqrt{2}(\bar{\beta}^2-1)-\bar{\beta}}\right)
e^{-3\alpha \bar{\beta}\sqrt{2}(\phi-\phi_*)}+\right. 
\]
\begin{equation}
\left.\frac{\bar{\beta}}{3\sqrt{2}(\bar{\beta}^2-1)-\bar{\beta}}
e^{-\alpha(\frac{3\sqrt{2}}{\beta}+1)(\phi-\phi_*)}\right]
\label{v}
\end{equation}
which shows that the assumption of constant equation of state in the $\phi$ dominated epoch requires a linear combination of two exponential terms for the potential.
\section{The parameterized equation of state}
Another point which should be stressed here is the fact that the cosmological equations for matter and energy dominated epochs are solvable even if the equation of state is not a constant. To see this, let us to assume that the equation of state varies with the scale factor. Given the parametrized function $\beta(N)$ (See \cite{parametrize} for some proposals for parametrizations of equation of state), it is possible to reconstruct the scalar field and it's potential  from equations (\ref{H})-(\ref{omegam}) and (\ref{omegaphi}). For matter dominated universe, using $H^2=H_*^2 X_m$ and (\ref{dotphi}), we have $\sqrt{ X_\phi}= \frac{\sqrt{2}\alpha}{\beta}\phi^\prime\sqrt{ X_m}$.  Substituting this in (\ref{omegam1}), yields:
\begin{equation}
\sqrt{ X_m}=\frac{1}{\sqrt2} e^{-(3N+\alpha(\phi-\phi_*))}
\label{mm}
\end{equation}
Therefore equation (\ref{omegaphi1}) can be expressed as:
\begin{equation}
4\alpha\phi^{\prime\prime}-2\alpha^2{\phi^\prime}^2+6\alpha(\beta^2-1)\phi^\prime-\beta^2=0
\label{38}
\end{equation}
In the matter dominated epoch, $ X_\phi\ll X_m$, and thus the second term is ignorable compared to the last term in  equation (\ref{38}). This makes it a linear second order differential equation for $\phi$ with the following solution:
\[
\phi(N)=\phi(0)+\phi\rq{}(0)\int^N_0 dN'\beta(N')e^{-3/2\int^{N'}_0dN''(\beta^2(N'')-1)}
\]
\begin{equation}
+\frac{1}{4\alpha}\int^N_0 dN'e^{-3/2\int^{N'}_0dN''(\beta^2(N'')-1)}\int^{N'}_0 dN''\beta^2(N'')e^{3/2\int^{N''}_0dN'''(\beta^2(N''')-1)}
\label{39}
\end{equation}
To reconstruct the field potential $V(\phi)$, using the relations (\ref{dotphi}), (\ref{vv}) and (\ref{mm}) we find that:
\begin{equation}
V=\frac{1}{2}(\frac{1}{\beta^2}-\frac{1}{2})\phi'^2H_*^2e^{-2(3N+\alpha(\phi-\phi_*))}
\end{equation}
Inverting $\phi(N)$ to $N(\phi)$ and substituting it into the above equation leads to the potential
$V(\phi)$.

A similar procedure can be carried out for dark energy dominated epoch. The result is:
\begin{equation}
\phi=\phi(0)\pm\frac{1}{\sqrt{2}\alpha}\int_0^N dN'\beta(N')
\end{equation}
\begin{equation}
V=\left( 1-\frac{\beta^2}{2}\right)\frac{H_*^2}{2\sqrt{2}\alpha^2}e^{-3/2\int_0^NdN' \beta^2(N')}
\left [ 
1+\frac{1}{2}\int_0^NdN'\beta(N')e^{\int_0^{N'} dN'' \left(3\beta^2(N'')/2-\beta(N'')/\sqrt{2}-3\right)}
\right ]
\label{42}
\end{equation}

As an application of equations (\ref{39})--(\ref{42}), we investigate the parameterization 
\begin{equation}
w=w_0+w_1\frac{z}{z+1}
 \end{equation}
 given in \cite{linder}. The best fit values are $w_0=0.05$ and $w_1=-0.55$. Equation (\ref{39}), for matter dominated epoch  leads to:
 \[
 \phi'=\frac{1}{12}\exp\left( -\frac{3Be^N}{2} -\left(\frac{3A}{2}+1\right)N\right)\left [ 12K +\frac{1}{\alpha}\left (\frac{-2}{3B}\right)^{3A/2}\right.
 \]
 \begin{equation}
 \left . \left ( -3(1+A)\Gamma\left(\frac{3A}{2},-\frac{3Be^N}{2}\right)+2\Gamma\left(1+\frac{3A}{2},-\frac{3Be^N}{2}\right)\right) \right ]
 \end{equation}
 where $K$ is an integration constant, $A=w_0+w_1=-0.5$ and $B=-w_1/(1+z^*)=-0.286$. 
  Using the expansion of incomplete gamma function $\Gamma(a,x)\simeq \Gamma(a)-x^a/a$ (for small $x$), we get the following form of $\phi$:
 \begin{equation}
 \phi(N)\simeq\left(2.12K-\frac{1}{6\alpha}\right)(-0.266+N)+\frac{1.47}{\alpha}+\frac{0.84}{\alpha}e^{0.75N}-0.67e^{0.429 e^N}+S
  \end{equation}
   where $S$ is another constant of integration. Both $K$ and $S$ can be eliminated in terms of $\phi(0)$ and $\phi'(0)$. In this  epoch we also have:
   \begin{equation}
   V\simeq\frac{1}{2}\left ( \frac{1}{1+A+Be^N}-\frac{1}{2}\right ) e^{3N-2\alpha(\phi-\phi(0))}\left(\frac{d\phi}{dN}\right)^2
   \end{equation}
   
   For dark energy dominated epoch similar calculations leads to:
   \begin{equation}
   \phi=\frac{\sqrt{2}}{\alpha}\left( \sqrt{0.5-0.286e^N}-0.7 \tanh^{-1}\left (1.4\sqrt{0.5-0.286e^N}\right)\right)+L
   \end{equation}
   where $L$ is an integration constant. Expanding terms in equation (\ref{42}) around $e^N=1$, one can show that   \begin{equation}
   V\simeq \frac{0.07 H_*^2}{\alpha^2}e^{-0.75 N}\left( 0.44 e^N-1\right)
   \end{equation}
\section{An exact solution}
\begin{figure}[htp]
\centering
\includegraphics[scale=0.6]{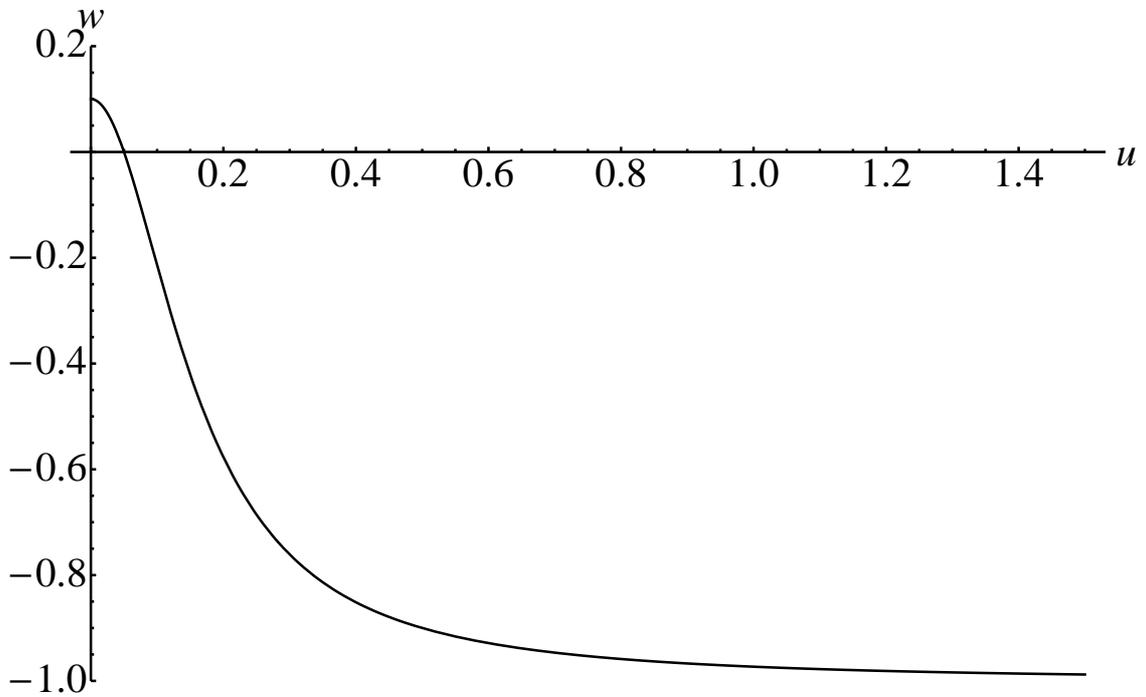}
\caption{Evolution of the equation of state with respect the normalized scale factor taking $c=5$, $\phi_0=0$, $\rho_m^0= 0.3\rho_c$ and $s=5.7\rho_c$ where $\rho_c$ is the cosmic critical density.}
\label{f7}
\end{figure}

\begin{figure}[htp]
\centering
\includegraphics[scale=0.6]{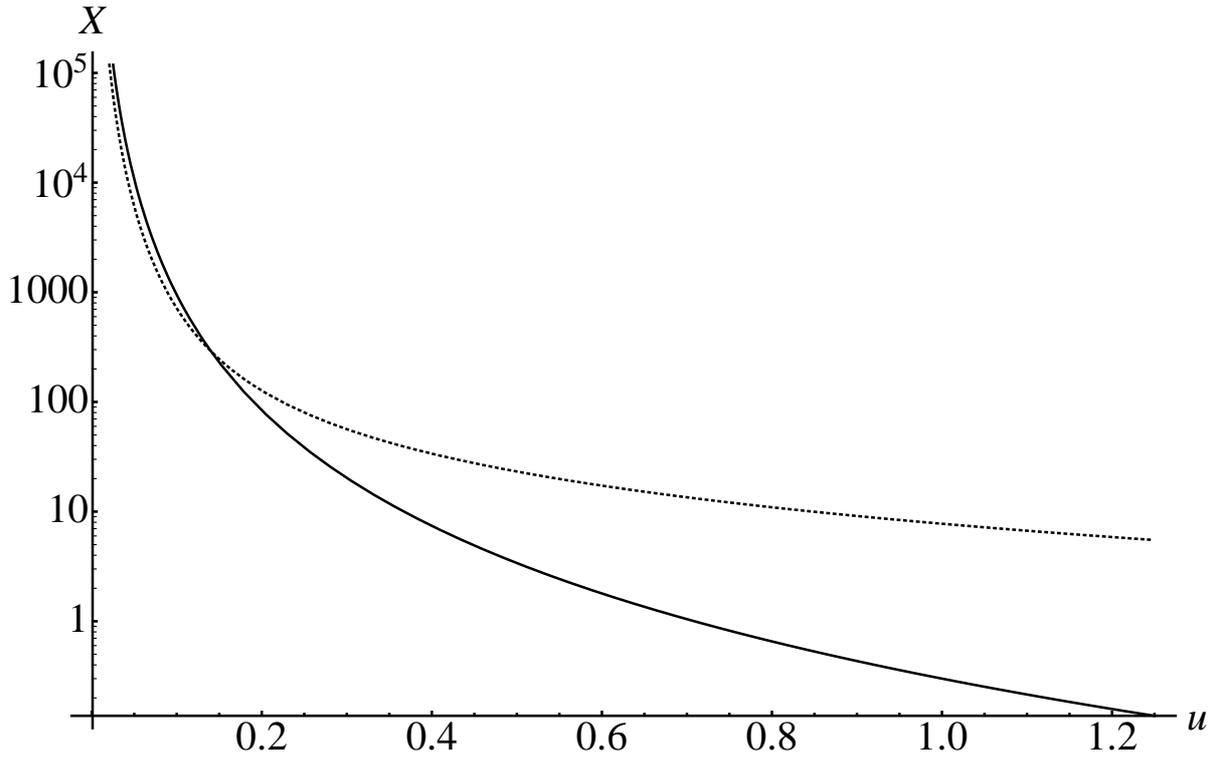}
\caption{Evolution of the density parameters taking $c=5$, $\phi_0=0$, $\rho_m^0= 0.3\rho_c$ and $s=5.7\rho_c$ }
\label{f8}
\end{figure}

Here we obtain some exact solution of equations (\ref{1})-(\ref{2})  without the assumption of constancy of the equation of state (For exact scaling solutions of this model with constant equation of state parameter see \cite{bat}). To do this let us rewrite the equations (\ref{1})-(\ref{2}) as follows:
\begin{equation}
H^2=2\alpha^2(\rho_m+\frac{1}{2}\dot{\phi}^2+V)
\label{11}
\end{equation}
\begin{equation}
\rho_m=\rho_m^0(\frac{a}{a_0})^{-3}e^{-\alpha(\phi-\phi_0)}
\label{rom}
\end{equation}
\begin{equation}
\ddot{\phi}+3H\dot{\phi}+\frac{dV}{d\phi}=\alpha\rho_m
\label{22}
\end{equation}
The second equation is derived by integrating equation (\ref{2}). Following the procedure of \cite{luis}, the above field equations  can be integrated to provide a variety of exact solutions of interacting quintessence valid for any epoch. Introducing $u=a/a_0$ and $dt=u^3d\eta$, equations (\ref{rom}) and (\ref{22}) leads to:
\begin{equation}
\frac{d^2\phi}{d\eta^2}+u^6\frac{dV}{d\phi}=\alpha \rho^0_m e^{\alpha\phi_0}u^3e^{-\alpha\phi}
\label{45p}
\end{equation}
 By writing $V=\frac{F(u)}{u^6}$, a first integral of equation (\ref{45p}) would be:
\begin{equation}
\frac{1}{2}\dot{\phi}^2+V=\frac{6}{u^{6}}\int du\frac{F(u)}{u} -\rho_m^0 u^{-3}e^{-\alpha(\phi-\phi_0)}+3\frac{\rho_m^0 e^{\alpha\phi_0}}{u^6} \int du u^2e^{-\alpha\phi}+\frac{E}{u^6}+D
\label{rophi}
\end{equation}  
Using this, one can express equation (\ref{11}) as:
\begin{equation}
H^2=2\alpha^2\left(\frac{6}{u^{6}}\int du\frac{F(u)}{u}+3\frac{\rho_m^0 e^{\alpha\phi_0}}{u^6}\int du u^2e^{-\alpha\phi}+\frac{E}{u^6}+D\right)
\end{equation} 
These are two coupled first order Integro--differential equations allowing  one to obtain $u(t)$ and $\phi(u)$ for a given function $F(u)$. In the above relation zero indices refer to the present values of quantities and $D$ and $E$ are 
arbitrary integration constants. Choosing $F(u)=nu^{r_1}+su^{r_2}$ and $E=D=0$, the corresponding exact solutions is:
\begin{equation}
\phi=c \ln{u}
\end{equation} 
\begin{equation}
V(\phi)=ne^{-\frac{(3+\alpha c)}{c}\phi}+ s e^{-6\alpha^2 c\phi}
\end{equation}
\[
\sqrt{2}\alpha(t-t_0)=\frac{\sqrt{1-\alpha^2 c^2}}{3\sqrt{s}\alpha^2 c^2}
\]
\[
\times\left[_2F_1\left(\frac{3\alpha^2 c^2}{6\alpha^2 c^2 -\alpha c -3},\frac{1}{2},1+\frac{3\alpha^2 c^2}{6\alpha^2 c^2 -\alpha c -3},\frac{3\rho_m^0 e^{\alpha \phi_0}(1-\alpha^2 c^2)}{(6\alpha^2 c^2 -\alpha c -3)s}u^{6\alpha^2 c^2 -\alpha c -3}\right)u^{3\alpha^2 c^2}\right.
\]
\begin{equation}
\left.-_2F_1\left(\frac{3\alpha^2 c^2}{6\alpha^2 c^2 -\alpha c -3},\frac{1}{2},1+\frac{3\alpha^2 c^2}{6\alpha^2 c^2 -\alpha c -3},\frac{3\rho_m^0 e^{-\alpha \phi_0}(1-\alpha^2 c^2)}{(6\alpha^2 c^2 -\alpha c -3)s}\right)\right]
\end{equation}
in which $s$ is an arbitrary integration constant, $n=\frac{\rho_m^0 e^{-\alpha \phi_0}\alpha c(1-3\alpha c)}{6\alpha^2 c^2 -\alpha c -3}$, $r_1=3-\alpha c$, $r_2=-6(\alpha^2 c^2-1)$ and we have assumed that $6\alpha^2 c^2-\alpha c-3\neq 0$. 
Using these relations, one can easily show that the dark energy and dark matter densities and the equation of state are respectively:
\begin{equation}
\rho_m=\rho_m^0 e^{\alpha \phi_0}u^{-(3+\alpha c)}
\end{equation}
\begin{equation}
\rho_\phi=\rho_m^0e^{\alpha \phi_0}\frac{\alpha c(1-6\alpha c)}{6\alpha^2 c^2-\alpha c-3}u^{-(3+\alpha c)}+\frac{s}{1-\alpha^2 c^2}u^{-{6\alpha^2 c^2}}
\end{equation}
\begin{equation}
\omega=\frac{\alpha\rho_m^0e^{\alpha \phi_0}(\alpha^2 c^2-1)-s(6\alpha^2 c^2-\alpha c-3)u^{-{6\alpha^2 c^2}+\alpha c +3}}{\alpha\rho_m^0e^{\alpha \phi_0}(\alpha^2 c^2-1)(6\alpha c-1)+s(6\alpha^2 c^2-\alpha c-3)u^{-{6\alpha^2 c^2}}}
\end{equation} 
Choosing the appropriate values for the parameters, $\alpha$ and $c$, in such a way that the the energy and matter densities are of the same order today, one can see that the solution is a tracker one. The  evolution of $\omega$ and the densities of each component are shown in figures (\ref{f7}) and (\ref{f8}). It can be seen that, by appropriate choosing of parameters, the equation of sate is a decreasing function with respect to the normalized scale factor. This is the main characteristic of  freezing models of quintessence \cite{cald}.  

\section{Conclusion}
In this paper we have investigated conformally coupled quintessence. To do this we have divided the evolution of the universe into matter and dark energy epochs. In any epoch the equation of state parameter of dark energy is assumed to be constant. We have found the appropriate range of this parameter to have a tracker behavior at matter dominated epoch. Moreover for a parametrized equation of state as a function of red shift, we have shown that this model leads to integrable equations in each epoch. Assuming a linear combination of two exponential terms for the potential,
an exact solution is presented which can be tracker by appropriate choosing of parameters.

\textbf{Acknowledgment} This work is  supported by a grant from university of Tehran. The authors would like to thank the anonymous referee for fruitful comments.


\end{document}